# Sequential Hybrid Beamforming Design for Multi-Link mmwave Communication

Yaning Zou[1], Mario Castañeda[2], Tommy Svensson[3] and Gerhard Fettweis[1]

[1] Vodafone Chair Technische Universität Dresden, Dresden, GERMANY
[2] Huawei Technologies Duesseldorf GmbH, Germany
[3] Chalmers University of Technology, Gothenburg, SWEDEN

Contact emails: {yaning.zou, fettweis}@ifn.et.tu-dresden.de, mario.castaneda@huawei.com, tommy.svensson@chalmers.se

*Abstract* —In this paper, we propose a sequential hybrid beamforming design for multi-link transmission over mmwave frequency bands. As a starting point, a baseline data communication link is established via traditional analog beamforming at both the BS and UE. If an extra RF chain is available at the UE, it can continue to probe the propagation environment at the same frequencies. In case the environment is favorable and system resources allow, a secondary data communication link is established to enable multi-stream transmission. In principle, the secondary link could be served by the same BS and/or one or several other BS(s). To initialize the secondary data communication link, a parallel beam search scheme is proposed, which helps the UE/BS to find a suitable beam pair with given optimization criteria without interrupting the baseline data communication. By applying the proposed two-step approach, hybrid beamforming becomes an add-on feature that can be easily switched on over an analog beamforming enabled system without interrupting its operation whenever system requires. Meanwhile, the information obtained by deploying the proposed parallel beam search scheme can also be used for deciding a back-up beam pair if signal blockage occurs to the baseline data communication link.

## I. INTRODUCTION

One of the most important aspects for the development of the 5th generation (5G) and future mobile communication network is to build cellular systems that can support extreme mobile broadband services, such as UHD/3D streaming, immersive applications and ultra-responsive cloud services [1]-[3]. In addition to optimize spectral efficiency and usage below 6 GHz, effective use of spectrum resources above 6 GHz for cellular communications has gradually been attracting extensive interests. Compared to its lower frequency counterpart, the biggest challenge for system design at the mmwave frequencies is to achieve good spectrum efficiency and at the same time combat significant path loss with reasonable cost [1], [4].

In this context, hybrid beamforming (HBF) is proposed to support spectrum efficient and flexible access with reasonable implementation cost and power consumption over mmwave frequency bands. In [4]-[7] and references therein, many hybrid beamforming techniques are proposed to design analog and digital beamformers with channel state information (CSI) knowledge of the complete propagation environment between base station (BS) and user equipment (UE). However, due to the use of large antenna arrays and inherently narrow beams, it generally requires long probe time to initialize reliable communication if the system needs to know the complete propagation channel. In addition, there is limited literature considering multi-node hybrid beamforming design which is a very important aspect in mmwave communication, due to the large blocking probability at those frequency bands.

In this paper, we propose a sequential hybrid beamforming design that is suitable for both single-node and multi-node single user transmissions. Assuming a baseline data communication link is first established using analog beamforming over the path with the strongest transmission power, then extra RF chain(s) at the UE can be used to continue probing the propagation environment. If the environment is favorable, a secondary data communication link could be established by connecting to the same BS and/or one or several other BS(s). To enable the proposed hybrid beamforming design, a parallel beam search scheme is proposed, which helps the UE/BS to find a suitable beam pair for the secondary data transmission without interrupting the baseline data communication.

## II. SIGNAL MODEL OF SEQUENTIAL HYBRID BEAMFORMING DESIGN FOR MULTI-LINK MMWAVE TRANSMISSION

In this paper, we consider single-user downlink (DL) transmission. Assume one UE can be served by one BS or $L$ BSs simultaneously. For simplicity of illustrating the basic concept, we assume the channel is frequency flat. At the $l$-th BS, $l \in \{1, 2, ..., L\}$, an $N_{BS,l}^h \times N_{BS,l}^v$ rectangular antenna array is implemented where $N_{BS,l}^h$ and $N_{BS,l}^v$ denote the numbers of antennas at the horizontal and vertical directions. At the serving UE, an $N_{UE} \times 1$ linear array is deployed where $N_{UE}$ refers to the number of antenna elements of the linear antenna array. We assume that $M_{BS,l}$ and $M_{UE}$ RF chains are available at the $l$-th BS and UE respectively. Based on the above-mentioned assumptions, a two-step sequential hybrid beamforming design is proposed as follows.

## A. Step 1: Analog beamforming for the baseline tranmission scheme

The first step of the proposed HBF design is to establish a baseline data transmission link using analog beamforming at the UE and at one BS. For notation simplicity, we name this BS as the BS #1. Any conventional procedure and algorithm developed for initializing such a link can be applied, see [4] and reference therein. Mathematically, the signal at the receiver detector input reads

$$\hat{s} = w_D \mathbf{w}_A^T \mathbf{H}_1 \mathbf{f}_A s + w_D \mathbf{w}_A^T \mathbf{n}, \quad (1)$$

where $s$ denotes the transmitted data with total transmit power of $\sigma_s^2 = E\left[|s|^2\right] = \sigma_X^2$, $\mathbf{f}_A$ refers to the $N_{BS,1}^h N_{BS,1}^v \times 1$ analog beamformer used at the BS, $\mathbf{w}_A$ and $w_D$ are the $N_{UE} \times 1$ analog beamformer and $1 \times 1$ equalization scalar applied at the UE, $\mathbf{n}$ is an $N_{UE} \times 1$ additive channel noise vector with complex Gaussian distributed entries following $CN\left(0, \sigma_n^2 \mathbf{I}\right)$. Furthermore, $\mathbf{H}_1$ refers to the $N_{UE} \times N_{BS,1}^h N_{BS,1}^v$ propagation channel from the BS #1 to the UE and can be modeled as in [7]-[9]. The detailed descriptions for constructing the analog beamformers as well as selecting a suitable analog beamforming pair can be found in [4] and reference therein.

## B. Step 2: Secondary data transmission using hybrid beamforming

Next, if the number of RF chains at the UE $M_{UE}$ is larger than 1, without interrupting the baseline data transmission, the UE can continue steering the antenna array with the second RF chain to find another serving path based on the given optimization criteria. In general, this serving link could be directed to the same BS (BS #1) or another BS (referred to as BS #2).

In case that the second link is served by the BS #1, at least 2 RF chains are required to be implemented at this BS. We denote the analog beam pair selected for activating this second link at the UE and BS #1 as $N_{UE} \times 1$ vector $\mathbf{w}'_{A,1}$ and $N_{BS,1}^h N_{BS,1}^v \times 1$ vector $\mathbf{f}'_{A,1}$, respectively. By applying two parallel analog beamformers at both the UE and BS #1, a $2 \times 2$ transmission link is established and the reception at the receiver detector input appears as

$$\hat{\mathbf{s}} = \mathbf{W}_{D,2}\left[\mathbf{w}_A \; \mathbf{w}'_{A,1}\right]^T \mathbf{H}_1\left[\mathbf{f}_A \; \mathbf{f}'_{A,1}\right]\mathbf{F}_{D,2}\mathbf{s} + \mathbf{W}_{D,2}\left[\mathbf{w}_A \; \mathbf{w}'_{A,1}\right]^T \mathbf{n}$$
$$= \mathbf{W}_{D,2}\mathbf{W}_{A+1}\mathbf{H}_1 \mathbf{F}_{A+1}\mathbf{x}_2 + \mathbf{W}_{D,2}\mathbf{W}_{A+1}\mathbf{n}, \quad (2)$$

where $\mathbf{W}_{A+1} = \left[\mathbf{w}_A \; \mathbf{w}'_{A,1}\right]^T$, $\mathbf{F}_{A+1} = \left[\mathbf{f}_A \; \mathbf{f}'_{A,1}\right]$, $\mathbf{s} = \left[s_1 \; s_2\right]$, $\mathbf{x}_2 = \mathbf{F}_{D,2}\mathbf{s}$, and $\mathbf{F}_{D,2}$ and $\mathbf{W}_{D,2}$ are $2 \times 2$ the digital precoding and combining matrices respectively. The total transmit power before feeding it into the transmit antennas is given by $\sigma_X^2 = E[trace\left(\mathbf{x}_2 \mathbf{x}_2^H\right)]$. The resulting $2 \times 2$ effective channel reads $\bar{\mathbf{H}}_2 = \mathbf{W}_{A+1}\mathbf{H}_1\mathbf{F}_{A+1}$ which enables the possibility of transmitting two independent streams. The digital beamformer $\mathbf{W}_{D,2}$ can be applied on the received signal for equalization using the effective channel matrix $\bar{\mathbf{H}}_2$ and digital precoder $\mathbf{F}_{D,2}$. This leads to a one-node HBF system.

On the other hand, if the second link is served by a second BS, i.e. BS #2, at least one free RF chain is required at the BS #2. Denoting the analog beam pair selected for activating the second link at the UE and BS #2 as $N_{UE} \times 1$ vector $\mathbf{w}''_{A,1}$ and $N_{BS,2}^h N_{BS,2}^v \times 1$ vector $\mathbf{f}''_{A,1}$, respectively and assuming perfect synchronization, we are able to establish multi-node transmission as

$$\hat{\mathbf{s}} = \mathbf{W}_{D,2}\left[\mathbf{w}_A \; \mathbf{w}''_{A,1}\right]^T \left(\mathbf{H}_1 \mathbf{f}_A \left[\mathbf{F}_{D,2}\mathbf{s}\right]_{1,1} + \mathbf{H}_2 \mathbf{f}''_{A,1}\left[\mathbf{F}_{D,2}\mathbf{s}\right]_{2,1}\right)$$
$$+ \mathbf{W}_{D,2}\left[\mathbf{w}_A \; \mathbf{w}''_{A,1}\right]^T \mathbf{n}$$
$$= \mathbf{W}_{D,2}\mathbf{W}'_{A+1}\left(\mathbf{H}_1 \mathbf{f}_A x_2^{1,1} + \mathbf{H}_2 \mathbf{f}''_{A,1}x_2^{2,1}\right) + \mathbf{W}_{D,2}\mathbf{W}'_{A+1}\mathbf{n},$$
$$\quad (3)$$

where $\mathbf{W}'_{A+1} = \left[\mathbf{w}_A \; \mathbf{w}''_{A,1}\right]^T$, $\mathbf{H}_2$ refers to the $N_{UE} \times N_{BS,2}^h N_{BS,2}^v$ propagation channel from the BS #2 to the UE and $x_2^{i,j}$ refers to the element at the $i$-th row and $j$-th column of $\mathbf{x}_2 = \mathbf{F}_{D,2}\mathbf{s}$ with digital precoder $\mathbf{F}_{D,2}$. In general, it is rather challenging to apply the digital precoding matrix $\mathbf{F}_{D,2}$ across different BSs and here we set $\mathbf{F}_{D,2} = \mathbf{I}$. The corresponding $2 \times 2$ effective transmission channel reads $\bar{\mathbf{H}}'_2 = \mathbf{W}'_{A+1}\left[\mathbf{H}_1 \mathbf{f}_A \; \mathbf{H}_2 \mathbf{f}''_{A,1}\right]$. Again, by applying digital beamforming $\mathbf{W}_{D,2}$ at the UE using the effective channel matrix $\bar{\mathbf{H}}'_2$, a 2-node HBF system is obtained.

Notice that, the signal models described in (2) and (3) are rather general, e.q. $s_1$ and $s_2$ are not necessary independent of each other. Thus, either a diversity or spatial multiplexing transmission scheme can be devised with proper digital beamformer design at the UE and BS(s). Meanwhile, as the single-user case is assumed in this paper, multi-user interference is not considered in the link-level models in (2) and (3), which forms a research topic for future study.

## C. Further step(s): Further secondary data transmission using hybrid beamforming

After the baseline and the secondary data transmission are established, the UE can still continue to probe the environment if one or more RF chains are available at the UE side. By repeating the same procedures as described in Section II. B, further secondary data transmission can be added one after another. The number of connecting secondary transmission links is determined by available system resources as well as performance targets. In this paper, we focus on connecting one secondary data transmission link in addition to the baseline data communication, which is the most likely case in practice due to limited implementation resources and size at the UE.

## III. BEAM SELECTION CRITERIA FOR THE PROPOSED TWO-STEP HYBRID BEAMFORMING DESIGN

In order to establish the secondary data transmission as in (2) and (3), the BS(s) and UE need to search a suitable beam pair $\{\mathbf{f}'_{A,1}(\mathbf{f}''_{A,1}), \mathbf{w}'_{A,1}\}$ in the beam space $\{\tilde{\mathbf{f}}^1_{k_1}(\tilde{\mathbf{f}}^2_{k_1}), \tilde{\mathbf{w}}_{k_2}\}$ where $k_1 = 1,...,K_{BS,l}$, $k_2 = 1,...,K_{UE}$, $K_{BS,l}$ and $K_{UE}$ denote the sizes of the codebooks at the $l$-th BS ($l = 1, 2$) and UE respectively. $\tilde{\mathbf{f}}^1_{k_1} (\tilde{\mathbf{f}}^2_{k_1})$ and $\tilde{\mathbf{w}}_{k_2}$ refer to the $k_1$-th and $k_2$-th entries of the codebooks implemented at the BS #1 (BS #2) and the UE respectively. Cascading with propagation channel and selected analog beamformers, the resulting test effective matrix reads $\tilde{\mathbf{H}}_{2,k_1,k_2} = [\mathbf{w}_A \ \tilde{\mathbf{w}}_{k_2}]^T \mathbf{H}_1 [\mathbf{f}_A \ \tilde{\mathbf{f}}^1_{k_1}]$ for the single-node scenario and $\tilde{\mathbf{H}}_{2,k_1,k_2} = [\mathbf{w}_A \ \tilde{\mathbf{w}}_{k_2}]^T [\mathbf{H}_1 \mathbf{f}_A \ \mathbf{H}_2 \tilde{\mathbf{f}}^2_{k_1}]$ for the 2-node scenario. If we further assume that the beam pair used in the baseline data transmission $\{\mathbf{f}_A, \mathbf{w}_A\}$ can not be chosen for the secondary data transmission, altogether $(K_{BS,1} - 1) \times (K_{UE} - 1)$ and $K_{BS,1}(K_{UE} - 1)$ beam pairs need to be examined if exhaustive search is deployed for the single-node and the multi-node scenarios. Then optimal beam pair entry indexes $\{k_{1,opt}, k_{2,opt}\}$ are selected by evaluating

$$\{k_{1,opt}, k_{2,opt}\} = \arg\max_{k_1, k_2} \Phi_{2,k_1,k_2}, \quad (4)$$

where the cost function $\Phi_{2,k_1,k_2}$ is designed based on desired performance target.

As one example, consider a DL transmission system that intends to increase the throughput by adding one independent transmission stream over its baseline data transmission. Assuming equal power for both streams at the BS(s) as $\sigma^2_{s_1} = \sigma^2_{s_2} = \sigma^2_X / 2$ and without precoding $\mathbf{F}_{D,2} = \mathbf{I}$, the cost function for finding suitable analog beam pairs to carry out one-node or two-node HBF for two stream spatial multiplexing based on (2) and (3) can be designed to maximize the achievable sum-rate of the effective channel or equivalently

$$\begin{aligned}\Phi_{2,RM,k_1,k_2} &= \det\left(\mathbf{I} + \frac{\sigma^2_X}{2\sigma^2_n} \mathbf{R}^{-1}_{2,k_1,k_2} \tilde{\mathbf{H}}^H_{2,k_1,k_2} \tilde{\mathbf{H}}_{2,k_1,k_2}\right) \\ &= \det\left(\rho \mathbf{R}^{-1}_{2,k_1,k_2}\right) \det\left(\tilde{\mathbf{H}}^H_{2,k_1,k_2} \tilde{\mathbf{H}}_{2,k_1,k_2} + \rho^{-1} \mathbf{R}_{2,k_1,k_2}\right)\end{aligned} \quad (5)$$

where $\mathbf{R}_{2,k_1,k_2} = \mathbf{W}^H_{A,k_2} \mathbf{W}_{A,k_2}$, $\rho = \sigma^2_{s_1} / \sigma^2_n = \sigma^2_{s_2} / \sigma^2_n = \sigma^2_X / (2\sigma^2_n)$ and $\mathbf{W}_{A,k_2} = [\mathbf{w}_A \ \tilde{\mathbf{w}}_{k_2}]$.

On other hand, the secondary communication link can also be designed to provide a diversity gain as well, e.g., by applying maximum ratio combining (MRC). In this paper, we focus on spatial multiplexing HBF design.

## IV. PROPOSED PARALLEL BEAM SEARCH SCHEME FOR ESTABLSHING SECONDARY DATA COMMUNICATION

One challenging aspect for establishing any communication link over mmwave bands is to initialize the data transmission with the least possible beam search time. In the proposed two-step sequential HBF design, data transmission is first established via analog beamforming. As only one path with the strongest transmission power needs to be found for conducting analog beamforming, the corresponding beam search time in principle should be much less than the probe time required for acquiring complete channel information required for deploying HBF designs in [6], [7]. Then after initialization has succeeded and data communication starts, we can search additional path(s) that supports additional data transmission from the same BS or another BS without interrupting the baseline data transmission *whenever* the system requires. Also, if the system detects that the channel condition for serving the baseline communication is weakening and sudden signal blockage appears, the beam search results for establishing the secondary communication link can be applied for selecting a back-up beam pair.

### A. Proposed Parallel Beam Training Frame Structure

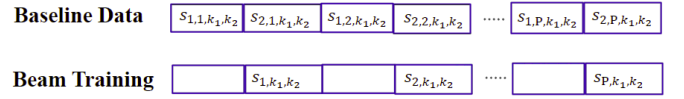

Fig. 1. Parallel beam search frame structure for test beam pair $\{\tilde{\mathbf{f}}^1_{k_1}(\tilde{\mathbf{f}}^2_{k_1}), \tilde{\mathbf{w}}_{k_2}\}$.

Based on (5), in order to evaluate the cost function for each test beam pair, the UE needs to know the effective test channel $\tilde{\mathbf{H}}_{2,k_1,k_2}$, the noise power $\sigma^2_n$ and the used hybrid beamformer $\mathbf{W}_{A,k_2}$ at the UE. Basically, we can assume that the latter two terms $\sigma^2_n$ and $\mathbf{W}_{A,k_2}$ are known at the UE. The main task is then to directly or indirectly evaluate the effective test channel.

Notice that, in case spatial multiplexing is deployed for multi-link single-node transmission and the transmission channel is unchanged after the beam search for analog beamforming, the BS #1 could already know the reception power for each individual path (the diagonal elements of $\tilde{\mathbf{H}}_{2,k_1,k_2}$) via analog beamforming in the first step. However this doesn't necessarily imply that BS #1 can directly make a good decision on selecting beam pair for establishing the secondary communication link. This is due to the fact that the cross-interference between the considered two paths (the off-diagonal elements of $\tilde{\mathbf{H}}_{2,k_1,k_2}$) are still unknown. Another round of beam search as proposed in this section is needed.

As shown in Fig. 1, an example parallel beam search frame structure is proposed. $S_{1,p,k_1,k_2}$ and $S_{2,p,k_1,k_2}$ are data symbols transmitted over the baseline link with identical transmission power $\sigma^2_s$. Assuming the data transmission continues as usual, the data transmission power reads $\sigma^2_s = \sigma^2_X$. $S_{p,k_1,k_2}$ refers to the $p$-th pilot symbol, $p = 1, \ldots, P$, sent from BS #1 for probing beam pair $\{\tilde{\mathbf{f}}^1_{k_1}, \tilde{\mathbf{w}}_{k_2}\}$ over the secondary test link in the single-node scenario. $S_{p,k_1,k_2}$ is transmitted from BS #2 for probing beam pair $\{\tilde{\mathbf{f}}^2_{k_1}, \tilde{\mathbf{w}}_{k_2}\}$ over the secondary test link in the 2-node scenario.

Notice that the detailed frame structure arrangement and pilot symbol number $P$ for each test beam pair can be flexibly modified and designed as long as the required observations can be obtained for carrying out the following estimation algorithms.

Now, assuming perfect synchronization, the $p$-th $2 \times 2$ data/pilot mixed matrix is transmitted and received using test beam pair $\{\tilde{\mathbf{f}}^1_{k_{1,1}}(\tilde{\mathbf{f}}^2_{k_{1,2}}), \tilde{\mathbf{w}}_{k_2}\}$ over two symbol durations. The corresponding observation at the UE becomes

$$\mathbf{y}_{p,k_1,k_2} = \tilde{\mathbf{H}}_{2,k_1,k_2} \bar{\mathbf{s}}_{p,k_1,k_2} + \mathbf{n}_{p,k_1,k_2}, \quad (6)$$

where $\mathbf{n}_{p,k_1,k_2}$ refers to a $2 \times 1$ additive channel noise vector after applying the test analog beamformer at the UE and

$$\overline{\mathbf{s}}_{p,k_1,k_2} = \begin{pmatrix} S_{1,p,k_1,k_2} & S_{2,p,k_1,k_2} \\ 0 & S_{p,k_1,k_2} \end{pmatrix}. \quad (7)$$

Based on (6), the training signal results in interference on the data transmission. Due to the sparsity of the mmwave propagation channel and directional nature of mmwave communication, the interference level is expected to be rather low for most test beam pairs. In this context, the power of the pilot transmission $\sigma_{sp}^2 = E\left[\left|S_{p,k_1,k_2}\right|^2\right]$ is also a design parameter and meanwhile is subject to a sum power constraint at the BS. With higher $\sigma_{sp}^2$, good estimation quality can be achieved with shorter pilot length $P$, yet the interference on the data transmission is larger. With lower $\sigma_{sp}^2$, good estimation quality can be achieved with longer pilot length $P$ and the interference on the data transmission is smaller. In case the interference level is non-negligible, more advanced signal processing algorithms should be deployed, which forms another research topic for future study.

For estimation purpose, two cases are considered in the following subsections. First, the data/pilot mixed matrix $\overline{\mathbf{s}}_{p,k_1,k_2}$ is assumed to be known if data symbols $s_{1,p,k_1,k_2}$ and $s_{2,p,k_1,k_2}$ are correctly demodulated. Second, $\overline{\mathbf{s}}_{p,k_1,k_2}$ is partially known or even totally unknown if no data demodulation is carried out before the beam probe process starts.

### A. Data/pilot matrix $\overline{\mathbf{s}}_{p,k_1,k_2}$ is known at the UE

Based on (6), if $\overline{\mathbf{s}}_{p,k_1,k_2}$ is known, the estimation of $\tilde{\mathbf{H}}_{2,k_1,k_2}$, i.e. the effective test channel for the baseline transmission and the trained beam pair for the secondary link is rather straightforward. For example, we apply least square (LS) estimation as

$$\hat{\tilde{\mathbf{H}}}_{2,p,k_1,k_2} = \mathbf{y}_{p,k_1,k_2}\overline{\mathbf{s}}_{p,k_1,k_2}^{-1}. \quad (8)$$

The quality of estimation can be improved by averaging over all $P$ estimations for each test beam pair. As the effective test channel is explicitly estimated here, it is possible to establish the secondary data communication immediately without further pilot training. However, this approach requires correct data demodulation under additive noise and interference from training symbol $s_{p,k_1,k_2}$.

### B. Data/pilot matrix $\overline{\mathbf{s}}_{p,k_1,k_2}$ is unknown at the UE

In case data symbols in (7) are not known, it is challenging to directly estimate the effective test matrix $\tilde{\mathbf{H}}_{2,k_1,k_2}$. On the other hand, the cost function in (5), not the effective test matrix $\tilde{\mathbf{H}}_{2,k_1,k_2}$, is the true estimation target. In this context, we propose two approaches to estimate or approximate the targeted cost function without carrying out any data demodulation.

For notation simplicity, we drop the $k_1$ and $k_2$ index in the continuation and define elements of the effective test channel $\tilde{\mathbf{H}}_{2,k_1,k_2}$, signal reception $\mathbf{y}_{p,k_1,k_2}$ and correlation matrix $\mathbf{R}_{2,k_1,k_2}$ as

$$\tilde{\mathbf{H}}_{2,k_1,k_2} \triangleq \begin{pmatrix} h_{11} & h_{12} \\ h_{21} & h_{22} \end{pmatrix} \quad (9a)$$

$$\mathbf{y}_{p,k_1,k_2} \triangleq \begin{pmatrix} y_{1,1,p} & y_{1,2,p} \\ y_{2,1,p} & y_{1,2,p} \end{pmatrix} + \mathbf{n}_p, \quad (9b)$$

where $y_{1,1,p} = S_{1,p,k_1,k_2}h_{1,1}$, $y_{1,2,p} = S_{2,p,k_1,k_2}h_{1,1} + S_{p,k_1,k_2}h_{1,2}$, $y_{2,1,p} = S_{1,p,k_1,k_2}h_{2,1}$, $y_{2,2,p} = S_{2,p,k_1,k_2}h_{2,1} + S_{p,k_1,k_2}h_{2,2}$ and

$$\mathbf{R}_{2,k_1,k_2} \triangleq \begin{pmatrix} r_{11} & r_{12} \\ r_{21} & r_{22} \end{pmatrix}. \quad (9c)$$

Incorporating (9a)-(9c) with (5), the second part of equation (5) is given by

$$\det\left(\tilde{\mathbf{H}}_{2,k_1,k_2}^H \tilde{\mathbf{H}}_{2,k_1,k_2} + \rho^{-1}\mathbf{R}_{2,k_1,k_2}\right) = G_1 + G_2, \quad (10a)$$

where

$$\begin{aligned} G_1 &= \left|h_{1,1}\right|^2\left|h_{2,2}\right|^2 + \left|h_{1,2}\right|^2\left|h_{2,1}\right|^2 - 2\operatorname{Re}\left[h_{1,1}h_{2,2}\left(h_{1,2}h_{2,1}\right)^*\right] \\ G_2 &= r_{1,1}\left(\left|h_{1,1}\right|^2 + \left|h_{2,1}\right|^2\right)/\rho + r_{2,2}\left(\left|h_{1,2}\right|^2 + \left|h_{2,2}\right|^2\right)/\rho \\ &\quad + r_{1,1}r_{2,2}/\rho^2 - r_{1,2}(h_{1,1}h_{1,2}^* + h_{2,2}h_{2,1}^*)/\rho \\ &\quad - r_{2,1}(h_{1,2}h_{1,1}^* + h_{2,1}h_{2,2}^*)/\rho - r_{1,2}r_{2,1}/\rho^2 \\ &\approx r_{1,1}\left(\left|h_{1,1}\right|^2 + \left|h_{2,1}\right|^2\right)/\rho + r_{2,2}\left(\left|h_{1,2}\right|^2 + \left|h_{2,2}\right|^2\right)/\rho \\ &\quad + r_{1,1}r_{2,2}/\rho^2 - r_{1,2}r_{2,1}/\rho^2. \end{aligned} \quad (10b)$$

The above approximation is made due to the fact that the off-diagonal elements of the correlation matrix $r_{1,2}$ and $r_{2,1}$ are much smaller than the diagonal elements $r_{1,1}$ and $r_{2,2}$ when reasonable large number of antenna elements are used for beamforming at the UE.

Assuming known $\sigma_n^2$ and $\mathbf{W}_{A,k_2}$ ($\mathbf{R}_{2,k_1,k_2}$), the values of $G_1$ and $G_2$ in (10b) can also be obtained by estimating $\left|h_{1,1}\right|^2$, $\left|h_{2,1}\right|^2$, $\left|h_{1,2}\right|^2$, $\left|h_{2,2}\right|^2$ and $h_{1,1}h_{2,2}\left(h_{1,2}h_{2,1}\right)^*$ instead of the effective test matrix directly. In details, based on the proposed pilot structure in (7), $\left|h_{1,1}\right|^2$, $\left|h_{2,1}\right|^2$, $\left|h_{1,2}\right|^2$ and $\left|h_{2,2}\right|^2$ are obtained by averaging over the $P$ observations with pilots as

$$\begin{aligned} \left|h_{1,1}\right|^2 &\approx \sum_{p=1}^{P}\left|y_{1,1,p}\right|^2/\left(P\sigma_s^2\right) = \left|h_{1,1}\right|^2\hat{\sigma}_{s_1}^2/\left(\sigma_s^2\right) \\ \left|h_{2,1}\right|^2 &\approx \sum_{p=1}^{P}\left|y_{2,1,p}\right|^2/\left(P\sigma_s^2\right) = \left|h_{2,1}\right|^2\hat{\sigma}_{s_1}^2/\left(\sigma_s^2\right) \\ \left|h_{1,2}\right|^2 &\approx \sum_{p=1}^{P}\left(\left|y_{1,2,p}\right|^2 - \left|y_{1,1,p}\right|^2\right)/\left(P\sigma_{sp}^2\right) \\ &= \left|h_{1,2}\right|^2\hat{\sigma}_{sp}^2/\sigma_{sp}^2 \\ &\quad + \underbrace{\left|h_{1,1}\right|^2\left(\hat{\sigma}_{s_2}^2 - \hat{\sigma}_{s_1}^2\right)/\left(\sigma_{sp}^2\right)}_{\approx 0} \\ &\quad + \underbrace{2\operatorname{Re}\left[h_{1,1}h_{1,2}^*S_{2,p,k_1,k_2}\left(S_{p,k_1,k_2}\right)^*\right]/\left(P\sigma_{sp}^2\right)}_{\approx 0} \end{aligned}$$

$$|h_{2,2}|^2 \approx \sum_{p=1}^{P}\left(\left|y_{2,2,p}\right|^2 - \left|y_{2,1,p}\right|^2\right)/\left(P\sigma_{sp}^2\right)$$
$$= |h_{2,2}|^2 \hat{\sigma}_{sp}^2 / \sigma_{sp}^2$$
$$+ \underbrace{|h_{2,1}|^2\left(\hat{\sigma}_{s_2}^2 - \hat{\sigma}_{s_1}^2\right)/\left(\sigma_{sp}^2\right)}_{\approx 0} \quad (11)$$
$$+ \underbrace{2\operatorname{Re}\left[h_{2,1}h_{2,2}^*S_{2,p,k_1,k_2}\left(S_{p,k_1,k_2}\right)^*\right]/\left(P\sigma_{sp}^2\right)}_{\approx 0}$$

where

$$\hat{\sigma}_{s_m}^2 = \sum_{p=1}^{P}\left|S_{m,p,k_1,k_2}\right|^2 / P$$
$$\hat{\sigma}_{sp}^2 = \sum_{p=1}^{P}\left|S_{p,k_1,k_2}\right|^2 / P,$$

and $m = 1,2$. For obtaining the approximation $I_1 \approx 0$ and $I_2 \approx 0$, we assume $P$ is sufficiently large so that $\left(\hat{\sigma}_{s_2}^2 - \hat{\sigma}_{s_1}^2\right)$ approaches to be zero, and $S_{2,p,k_1,k_2}$ and $S_{p,k_1,k_2}$ are uncorrelated.

Next, we set $S_{p,k_1,k_2} = \sqrt{\beta}S_{1,p,k_1,k_2}^*$, $\beta = \sigma_s^2 / \sigma_{sp}^2$ and evaluate the cross term as $h_{1,1}h_{2,2}\left(h_{1,2}h_{2,1}\right)^*$ based on

$$h_{1,1}h_{2,2} \approx \sum_{p=1}^{P} y_{1,1,p}y_{2,2,p} / \left(P\sigma_{sp}\sigma_s\right)$$
$$= h_{1,1}h_{2,2}\sum_{p=1}^{P}\left|S_{1,p,k_1,k_2}\right|^2 / \left(P\sigma_{sp}\sigma_s\right)$$
$$+ \underbrace{h_{2,1}h_{1,1}\sum_{p=1}^{P}S_{1,p,k_1,k_2}S_{2,p,k_1,k_2} / \left(P\sigma_{sp}\sigma_s\right)}_{\approx 0}$$
$$\qquad (12)$$
$$h_{1,2}h_{2,1} \approx \sum_{p=1}^{P} y_{1,2,p}y_{2,1,p} / \left(P\sigma_{sp}\sigma_s\right)$$
$$= h_{1,2}h_{2,1}\sum_{p=1}^{P}\left|S_{1,p,k_1,k_2}\right|^2 / \left(P\sigma_{sp}\sigma_s\right)$$
$$+ \underbrace{h_{1,1}h_{2,1}\sum_{p=1}^{P}S_{1,p,k_1,k_2}S_{2,p,k_1,k_2} / \left(P\sigma_{sp}\sigma_s\right)}_{\approx 0}.$$

In (12), we assume $S_{1,p,k_1,k_2}$ and $S_{2,p,k_1,k_2}^*$ are uncorrelated and $P$ is large enough so that $\sum_{p=1}^{P}S_{1,p,k_1,k_2}S_{2,p,k_1,k_2} / P$ approaches to zero.

It is interesting to notice that no signal demodulation for the data symbols or pilot symbols is required in (11) and (12). The formulas in (11) focus on estimations of amplitude information of the effective test channel while the formulas in (12) estimate both amplitude and phase information of the effective test channel. With limited pilot length $P$ and/or pilot transmission power $\sigma_{sp}^2$, accurate phase estimation could be very challenging.

In this context, we seek an evaluation approach that uses only the amplitude information of the received observations. In details, we approximate the real part of the cross term $\operatorname{Re}\left[h_{1,1}h_{2,2}\left(h_{1,2}h_{2,1}\right)^*\right]$ by its amplitude $|h_{1,1}||h_{2,2}||h_{1,2}||h_{2,1}|$. Inherently, the evaluation of $G_1$ becomes

$$G_1 \approx |h_{1,1}|^2|h_{2,2}|^2 + |h_{1,2}|^2|h_{2,1}|^2 - 2\sqrt{|h_{1,1}|^2|h_{2,2}|^2|h_{1,2}|^2|h_{2,1}|^2}.$$
$$\qquad (13)$$

To evaluate the above approximation in (13), it requires only estimation of the power of each element in the test effective matrix as in (11) and doesn't need a special probe pilot symbol design as long as the pilot symbol $S_{p,k_1,k_2}$ is uncorrelated with the transmitted data symbols.

As will be shown later in the simulations, this approximation in (13) can provide reasonably good indications for choosing a suitable beam pair and outperforms the beam search algorithm based on estimation of $G_1$ in (10b)-(12) with a small amount of system overhead.

## V. NUMERICAL RESULTS AND ANALYSIS

### A. System Setups

In this section, we examine the proposed two-step HBF design using extensive computer simulations. The transmitted power is 27 dBm and the distance between the BS #1 and UE is set to be 30 m. In both the single-node and 2-node cases, equal transmission power per stream is assumed. The overall channel power between the BS #2 and the UE is minus infinite to 6dB less than that between the BS #1 and UE. The QuaDRiGa geometry-based 3D stochastic channel simulator is deployed for generating the channel propagation environment at 28 GHz based on measurement results obtained in the mmMAGIC project [9], [10]. An $8 \times 8$ planar antenna array and a $1 \times 8$ uniform linear array are implemented at the BS and the UE sides respectively. The codebook entries are assumed to be equally distributed over the steering angle domain, i.e., $[-\pi, \pi)$ for azimuth angle and $[-\pi/2, \pi/2)$ for elevation angle based on given codebook sizes $K_{BS,l} = K_{BS,l,A} \times K_{BS,l,E}$, $l = 1,2$, at the BS side and $K_{UE} = K_{UE,A} \times 1$ at the UE side. Here $K_{BS,l,A}$ ($K_{UE,A}$) denotes the number of codebook entries at the azimuth angle and $K_{BS,l,E}$ represents the number of codebook entries at the elevation angle, and we select $K_{BS,l,A} = 32$, $K_{BS,l,E} = K_{UE,A} = 16$ and $P = 10$. As one numerical example, assuming sampling rate at 122.88 MHz, the exhaustive search requires a 0.66 ms probe time period.

In Figure 2-3, achievable sum-rates of the proposed two-step sequential HBF design are evaluated in both the single-node and the 2-node scenarios. The optimal HBF design using perfect CSI for the single-node scenario [7] is deployed for comparison purpose. As shown in Fig. 3, there is about 0.5 dB performance loss for using the proposed beamforming design compared to the optimal HBF with two stream transmission. This is due to the fact that the proposed approach uses CSI of the effective channel that is limited by the used codebook sizes while the optimal approach uses perfect CSI of the true propagation channel. In both single-node and multi-node scenarios, with given design parameters, the estimation methods developed based on (8) (PHBF #1) as well as based on (10a)-(11), (13) (PHBF #3) can effectively provide performance close to the case using perfect CSI of the effective channel. In general, the proposed hybrid beam search method (PHBF #3) appears to be the most efficient estimation approach with limited $P$.

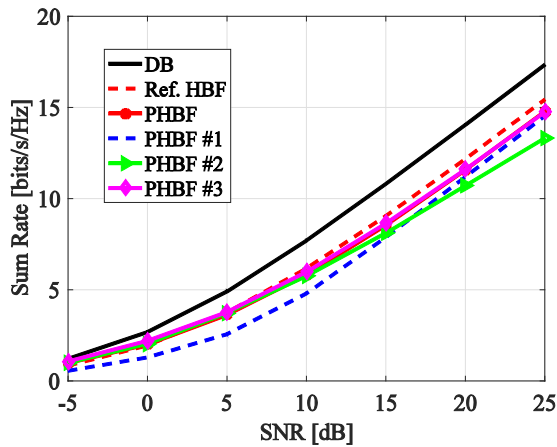

Fig. 2. Performance comparisons of digital beamforming (DB), reference HBF, the proposed HBF using perfect CSI of the effective channel (PHBF), estimated CSI as in (8) (PHBF #1), approximation based on (10a)-(12) (PHBF #2) and approximation based on (10a)-(11), (13) (PHBF #3). Single-node scenario. SNR refers to received SNR at each UE antenna input.

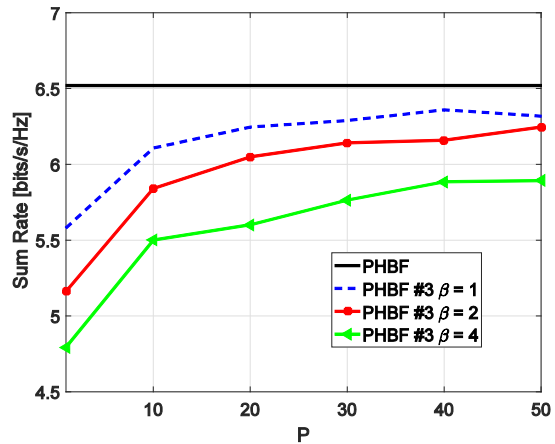

Fig. 4. Performance comparisons of the proposed HBF using estimated beam search cost function in (10) based on (13) (PHBF #3) with perfect CSI knowledge (PHBF) and different pilot transmission powers (PHBF #3, $\beta = 1, 2, 4$, $\beta = \sigma_s^2 / \sigma_{sp}^2$). SNR=10 dB. Single-node scenario. SNR refers to received SNR at each UE antenna input.

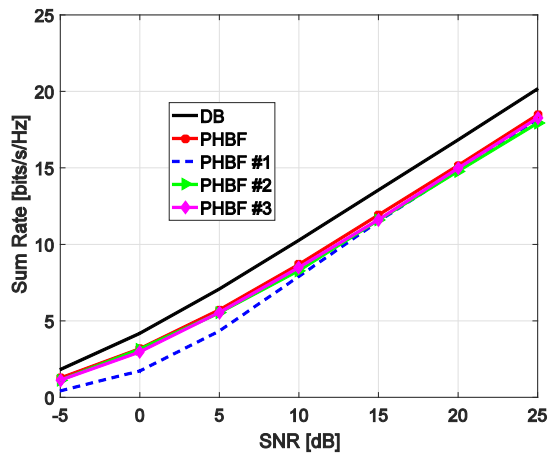

Fig. 3. Performance comparisons of digital beamforming (DB), reference HBF, the proposed HBF using perfect CSI of the effective channel (PHBF), estimated CSI as in (8) (PHBF #1), estimated beam search cost function in (10) (PHBF #2) and estimated beam search cost function in (10) based on (13) (PHBF #3). 2-node scenario. SNR refers to received SNR at each UE antenna input from BS #1.

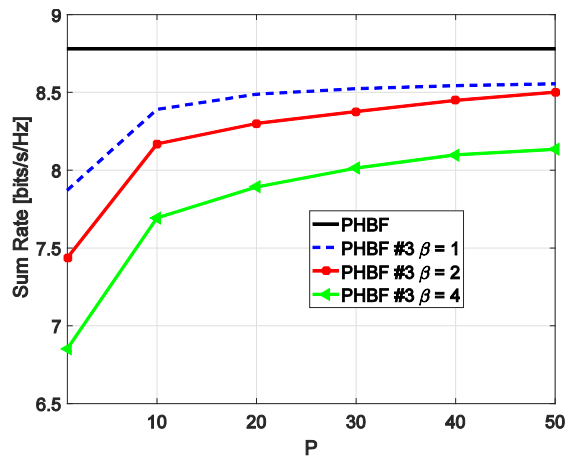

Fig. 5. Performance comparisons of the proposed HBF using estimated beam search cost function in (10) based on (13) (PHBF #3) with perfect CSI knowledge (PHBF) and different pilot transmission powers (PHBF #3, $\beta = 1, 2, 4$, $\beta = \sigma_s^2 / \sigma_{sp}^2$). SNR=10 dB. Two-node scenario. SNR refers to received SNR at each UE antenna input from BS #1.

Next the impact of the design parameters is evaluated using PHBF #3. As shown in Fig. 4-5, we compare the achievable sum rates using three different pilot transmission power levels and pilot symbol length $P$. Denoting $\beta = \sigma_s^2 / \sigma_{sp}^2$, there is a clear trade-off between the pilot transmission power and pilot symbol length. As shown in Fig. 6-7, the cumulative density function (CDF) of the signal-to-noise-plus-interference ratio (SINR) of the baseline data transmission under the proposed parallel beam training scheme is compared with different $\beta$ values. It shows that parallel beam training has little impact on the baseline data transmission in the SNR range of interest for mmwave communication.

## VI. CONCLUSIONS

In this paper, we proposed a two-step sequential hybrid beamforming design that can be deployed for multi-link mmwave transmission. The basic idea is to minimize the hybrid beamforming initialization time by first deploying analog beamforming over the strongest path. Then BS(s) and UE can continue to search path(s) for establishing multi-stream transmission using extra RF chains based on the desired performance target whenever the system requires. Due to the channel sparsity and directional nature of mmwave communication, the training signal can be sent without interrupting baseline data transmission. Further development of the scheme will be carried out to cope with the case when the interference level caused by the pilot transmission is non-negligible and for frequency-selective channels.

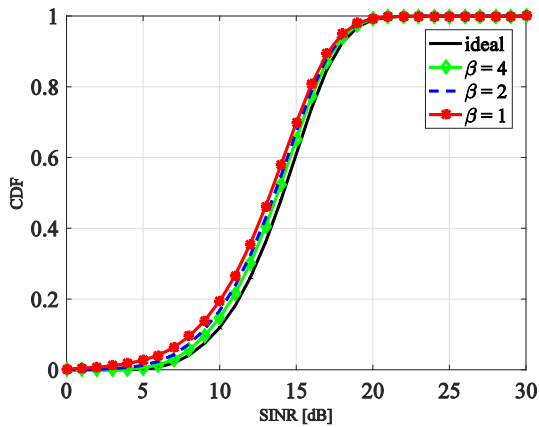

Fig. 6. CDF of SINR on the baseline data transmission when parallel beam training is carried out with different pilot transmission powers. $\beta = \sigma_s^2 / \sigma_{sp}^2$. SNR=10 dB. Single-node scenario. SNR refers to received SNR at each UE antenna input.

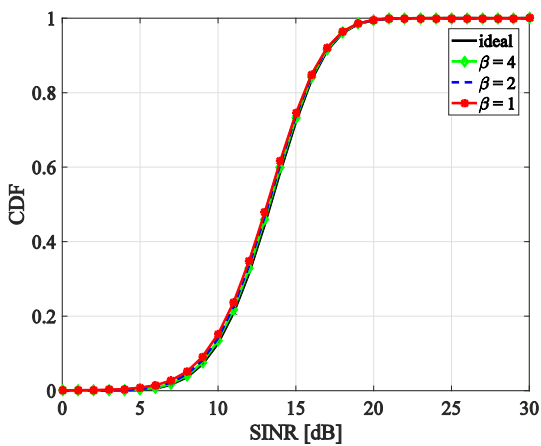

Fig. 7. CDF of SINR on the baseline data transmission when parallel beam training is carried out with different pilot transmission powers. $\beta = \sigma_s^2 / \sigma_{sp}^2$. SNR=10 dB. 2-node scenario. SNR refers to received SNR at each UE antenna input from BS #1.

## VII. ACKNOWLEDGEMENT

The research leading to these results received funding from the European Commission H2020 programme under grant agreement n°671650 (5G PPP mmMAGIC project).